\documentclass[journal]{IEEEtran}

\usepackage{url}

\usepackage{todonotes}

\usepackage{graphicx}               % Necessary to use \scalebox

\ifCLASSINFOpdf
  \graphicspath{{img/}}
  \DeclareGraphicsExtensions{.pdf,.jpeg,.png}
\else
\fi
\usepackage{amsmath}
\begin{document}
\title{Analysis of skin lesion images with deep learning}

\author{Josef~Steppan,
        Sten~Hanke% <-this % stops a space
\thanks{J. Steppan was with the Department
of eHealth at FH Joanneum University of Applied Sciences, Alte Poststrasse 149, 8020 Graz, AUSTRIA. e-mail: josef.steppan@edu.fh-joanneum.at}% %<-this % stops a space
\thanks{S. Hanke is with the Department
of eHealth at FH Joanneum University of Applied Sciences, Alte Poststrasse 149, 8020 Graz. e-mail: sten.hanke@fh-joanneum.at}% <-this % stops a %space
}
\maketitle

\begin{abstract}
Skin cancer is the most common cancer worldwide, with melanoma being the deadliest form. Dermoscopy is a skin imaging modality that has shown an improvement in the diagnosis of skin cancer compared to visual examination without support. We evaluate the current state of the art in the classification of dermoscopic images based on the ISIC-2019 Challenge for the classification of skin lesions and current literature. Various deep neural network architectures pre-trained on the ImageNet data set are adapted to a combined training data set comprised of publicly available dermoscopic and clinical images of skin lesions using transfer learning and model fine-tuning. The performance and applicability of these models for the detection of eight classes of skin lesions are examined. Real-time data augmentation, which uses random rotation, translation, shear, and zoom within specified bounds is used to increase the number of available training samples. Model predictions are multiplied by inverse class frequencies and normalized to better approximate actual probability distributions. Overall prediction accuracy is further increased by using the arithmetic mean of the predictions of several independently trained models. The best single model has been published as a web service. The source code is publicly available at \url{http://github.com/j05t/lesion-analysis}
\end{abstract}


% \fi
\IEEEpeerreviewmaketitle

\section{Introduction}
\IEEEPARstart{S}{kin} cancer is the most common cancer worldwide, with melanoma being the deadliest form.
A later stage in the diagnosis of melanoma is associated with a strong influence on melanoma mortality within 5 years of diagnosis \cite{wernli2016screening}. Early detection of melanoma can significantly reduce both morbidity and mortality \cite{di2018computer}. The risk of dying from the disease is directly related to the depth of the cancer, which is directly related to the time it has been growing. Self-examination of the skin by patients, full-body skin examinations by a doctor, and patient education are the keys to early detection. Self-examiners are generally diagnosed with thinner melanomas than non-self-examiners (0.77 mm versus 0.95 mm) \cite{carli2003dermatologist}.

%Visual skin cancer screening has been shown to reduce melanoma-associated mortality rates (Figure ~\ref{mortality}) \cite{wernli2016screening}.

This paper evaluates the current state of the art in the classification of dermoscopic images based on the ISIC-2019 Challenge for the classification of skin lesions and current literature. Since medical image data sets often show a class imbalance, several approaches for the training of deep neural networks on imbalanced data sets have been reviewed. Because the training of deep neural networks requires a large amount of training data, further publicly available dermoscopic as well as clinical image data sets of skin lesions have been evaluated for expanding the ISIC-2019 training data set. Since the heterogeneity of the image data of the ISIC data set requires preprocessing, a suitable approach towards preprocessing, as well as the effects of preprocessing on the achieved accuracy of trained networks have been investigated. Furthermore, the potential of real-time data augmentation to increase the number of available training patterns during training and to improve the prediction accuracy at inference time has been investigated. Current ensembling strategies and an overview of current architectures of deep neural networks for the classification of image content have been reviewed.

%\begin{figure}[!t]
%\centering
%\includegraphics[width=3.49 in]{tablle.png}
%\caption{Melanoma Mortality Associated With Visual Skin Cancer Screening: Screening for Skin Cancer in Adults: An Updated Systematic Evidence Review for the U.S. Preventive Services Task Force [Internet]. Evidence Syntheses, No. 137. \cite{wernli2016screening}}
%\label{mortality}
%\end{figure}

%\iffalse

%\subsection{Screening}
%\todo{here some extension}
%\fi

%\IEEEpubidadjcol
\section{Image Classification}
%\subsection{State of the art}
Convolutional Neural Networks (CNNs) \cite{krizhevsky2012imagenet} are currently state of the art in image classification and have been exceeding the recognition rate of human experts in the ImageNet Large Scale Visual Recognition Challenge\footnote{\url{http://image-net.org/challenges/LSVRC/2017}} (ILSVRC) \cite{russakovsky2015imagenet} since 2015 \cite{Langlotz}.
%(Figure~\ref{history_error_rates}.
The ILSVRC evaluates algorithms for object recognition and image classification on a large scale. An important motivation is to enable researchers to compare progress in recognition for a wider variety of objects. Another motivation is to measure the progress of computer vision algorithms for classifying images on a large scale. The ImageNet training data set contains 1.000 categories and 1,2 million images. Image classification algorithms are compared using a test data set of 150.000 images in 1.000 categories. Highest accuracy rates are currently achieved with the architectures SENet \cite{hu2018squeeze} 154 (81.3\% top-1 accuracy), PNASNet-5 Large \cite{DBLP:journals/corr/abs-1712-00559} (82.9\%), AmoebaNet-C \cite{Liu_2018_ECCV, real2019regularized} (83.9\%) and EfficientNet-B7 \cite{ DBLP:journals/corr/abs-1905-11946} (84.4\%) \cite{bianco2018benchmark}.
Algorithms for classifying image content are constantly being improved. Deep learning has shown enormous potential in this area due to the constantly increasing amounts of data \cite{cui2019assessing,  fujisawa2019possibility}. Some deep learning approaches outperform teams of certified dermatologists in the detection of melanoma in dermoscopic images \cite{hekler2019superior,  maron2019systematic, brinker2019deep} or achieve equivalent detection rates \cite{ blum2004digital, zortea2014performance}.

%\begin{figure}[!t]
%\centering
%\includegraphics[width=3.0 in]{comparison.png}
%\caption{History of error rates for image classification at the ImageNet large-scale visual recognition challenge. Accuracy improved dramatically with the introduction of Deep Learning in 2012 and continued to improve afterwards. Humans achieve error rates of about 5\% (\cite{Langlotz})}
%\label{history_error_rates}
%\end{figure}

%\begin{figure}[ht]
%\centering
%\includegraphics[width=3.49 in]{accuracy.png}
%\caption{Ball diagram displaying top-1 accuracies on ImageNet data and computational complexity for several deep neural network architectures for image classification. Only the FLOPs (Center Crop versus Floating Point Operations) required for a single forward pass are shown. The
%size of the circle corresponds to the complexity of the model. Graphic from "Benchmark analysis of representative deep neural network architectures \cite{bianco2018benchmark}}
%\label{ball_diagram_accuracy}
%\end{figure}

%\begin{figure}[ht]
%\centering
%\includegraphics[width=3.49 in]{flops_efficientnet.png}
%\caption{EfficientNet: FLOPs vs top-1 accuracy on ImageNet. Plot from Efficientnet: Rethinking model scaling for convolutional neural networks,  \cite{DBLP:journals/corr/abs-1905-11946}}
%\label{flops_efficientnet}
%\end{figure}

%\subsection{Image Classification}

\section{Skin Lesion Datasets} \label{section_datasets}

\subsection{ISIC-2019}
To make specialist knowledge more widely available, the International Skin Imaging Collaboration developed the ISIC archive, an international repository for dermoscopic images, both for clinical training purposes and to support technical research on automated algorithmic analysis by hosting the ISIC Challenges. The training data set of the ISIC-2019 Challenge consists of several dermoscopic image databases: BCN\_20000 \cite{combalia2019bcn20000} with dermoscopic images of the most common classes of skin lesions: actinic keratosis, squamous cell carcinoma, basal cell carcinoma, seborrheic keratosis, solar lentigo, and dermatological lesions. The HAM10000 dataset \cite{tschandl2018ham10000}, with 600x450 images centered and cropped on lesions. The MSK data set \cite{codella2018skin} with images of different resolutions. A total of 25,331 images are available for training in 8 different categories. The test data set consists of 8,238 images whose labels are not publicly available. Also, the test data set contains an additional outlier class that is not contained in the training data and must be identified by developed systems. Predictions on the ISIC-2019 test data set are assessed by an automatic evaluation system. The goal of the ISIC-2019 Challenge\footnote{\url{https://challenge2019.isic-archive.com/}} is to classify dermoscopic images among nine different diagnostic categories:
\begin{enumerate}
	\item  Melanoma (MEL)
	\item  Melanocytic nevus (NV)
	\item  Basal cell carcinoma (BCC)
	\item  Actinic Keratosis (AK)
	\item  Benign keratosis (solar lentigo / seborrheic keratosis / lichen planus-like keratosis) (BKL)
	\item  Dermatofibroma (DF)
	\item  Vascular Lesion (VASC)
	\item  Squamous cell carcinoma (SCC)
	\item  None of the others (UNK)
\end{enumerate}

\subsection{PH2 database}
The PH2 database \cite{mendoncca2013ph} includes manual segmentation, clinical diagnosis, and the identification of multiple dermoscopic structures performed by experienced dermatologists in a set of 200 dermoscopic images. The images were obtained in the dermatology department of the Pedro Hispano hospital (Matosinhos, Portugal) under the same conditions by the Tuebinger Mole Analyzer System using 20-fold magnification. These are 8-bit RGB color images with a resolution of 768x560 pixels. The image database contains a total of 200 dermoscopic images of melanocytic lesions, including 80 common nevi, 80 atypical nevi, and 40 melanomas. The PH2 database contains a medical annotation of all images, namely a medical segmentation of the lesion, a clinical and histological diagnosis as well as the evaluation of several dermoscopic criteria (colors; pigment network; dots/spheres; stripes; regression areas; blue-whitish haze). The database was made freely available for research and benchmarking purposes\footnote{https://www.fc.up.pt/addi/ph2\%20database.html}.

\subsection{Light Field Image Dataset of Skin Lesions}
Faria et al. \cite{de2019light} present a contribution to the research community in the form of the publicly available data set of skin lesions, the "Light Field Image Dataset of Skin Lesions" (SKINL2)\footnote{https://www.it.pt/AutomaticPage?id=3459}. The dataset contains 250 light fields \cite{wu2017light}, which were recorded with a focused plenoptic camera and divided into eight clinical categories depending on the type of lesion. Each light field consists of 81 different views of the same lesion. The database also contains the dermoscopic image of each lesion. The data set offers great potential the further development of medical imaging research and the development of new classification algorithms based on light fields as well as for clinically oriented dermatological studies; however, only dermoscopic images contained in the data set are taken into account for this work.

\subsection{SD-198} \label{subsec_sd198}
In contrast to dermoscopic images with largely constant lighting and low image disturbances, clinical images are often created with a large number of different image recording devices, such as digital cameras or smartphones. The SD-198 data set \cite{sun2016benchmark} contains 6,584 clinical images from 198 classes, which vary according to scale, color, shape, and structure. The SD-198 benchmark data set is intended to stimulate further research into the visual classification of skin diseases. The authors also carry out an extensive analysis of this data set using modern methods including CNNs. The ground truth labels of the images were created via DermQuest\footnote{\url{https://www.dermquest.com}}, with each image being examined by qualified experts and labeled with the name of its class. To ensure the quality of the labels, two experts were also invited to check the data set.

\subsection{7-point criteria evaluation database}
Kawahara et al. \cite{Kawahara2019-7pt} provide a database for evaluating the computerized image-based prediction of the 7-point checklist for malignant skin lesions\footnote{\url{https://derm.cs.sfu.ca/Welcome.html}}. The seven-point checklist, published in 1998, is one of the best-validated dermoscopic algorithms due to its high sensitivity and specificity, even when used by non-specialists. The seven criteria were originally applied to 342 melanocytic lesions (117 melanomas and 225 atypical nevi) tested and selected for their frequent association with melanoma \cite{argenziano1998epiluminescence}. Three of them were defined as the main criteria (atypical network, blue-white haze, and atypical vascular pattern), while the remaining four were considered minor (irregular stripes, irregular spots or globules, irregular spots, and regression structures) \cite{kittler2016standardization}. The data set contains over 2000 clinical and dermoscopic color images as well as corresponding structured metadata that are tailored to the training and evaluation of CAD (Computer Aided Diagnostic) systems.

\subsection{MED-NODE}
The MED-NODE data set \cite{giotis2015med} consists of 70 melanoma and 100 nevus images from the digital image archive of the Department of Dermatology at the University Hospital Groningen (UMCG), which is used for the development and testing of the MED-NODE Decision Support System for the detection of Skin cancer using macroscopic images. The system proposed by the authors achieves results with a diagnostic accuracy of 81\%. The final classification was achieved by a majority vote of the predictions of several models. The dataset is publicly available\footnote{\url{http://www.cs.rug.nl/~imaging/databases/melanoma\_naevi/}}.

\section{Combined training dataset}
\begin{figure}[!t]
\centering
\includegraphics[width=3.49in]{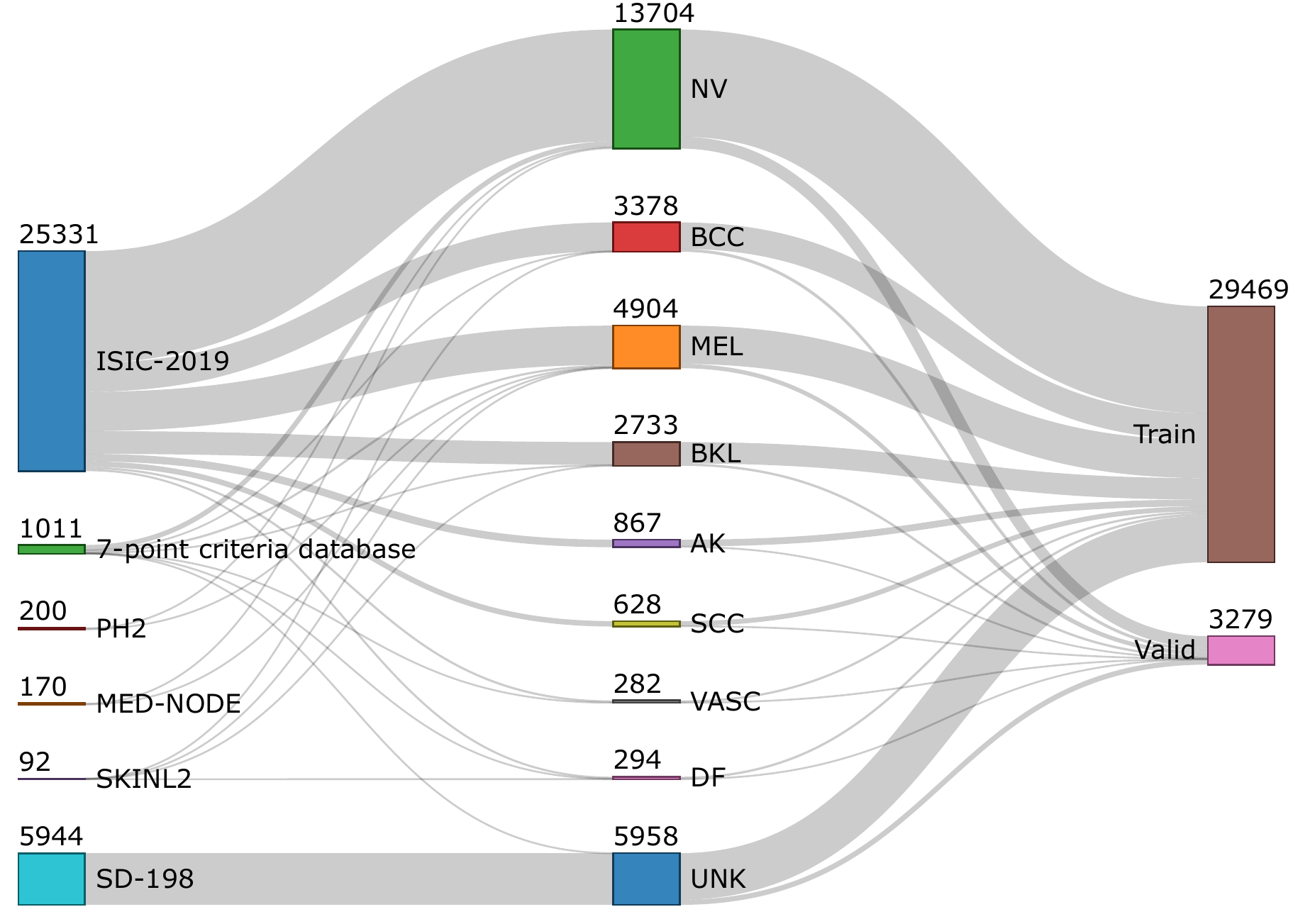}
\caption{Combined training data set from the data sets ISIC-2019, PH2, Light Field Image Dataset of Skin Lesions, SD-198, the 7-point criteria evaluation database, and MED-NODE. The "UNK" category is mainly formed from data from the SD-198 dataset. The combined data set is divided into a training (90\%) and validation data set (10\%), so 29.469 images are available for training and 3.279 images for assessing the generalizability of the predictions and for adapting hyperparameters in the validation data set. The ISIC-2019 test data set consists of 8.238 images whose labels are not publicly available. The test data set is not used for training or parameter adjustment.}
\label{dataset}
\end{figure}

A combined training data set has been created from all the data sets described in section ~\ref{section_datasets}. 32,748 images are available for training in total. Images from SD-198 were used exclusively for the creation of training data for the "UNK" class, after prior removal of image data from the eight categories of the ISIC-2019 training data set. The combined data set is still heavily imbalanced (Figure ~\ref{dataset}).

\section{Methodology}

\subsection{Preprocessing}
\begin{figure}[ht]
\centering
\includegraphics[width=3.2 in]{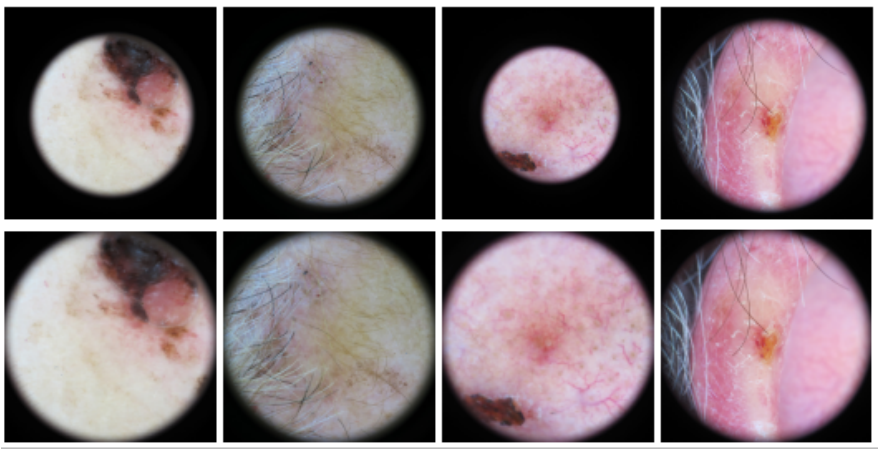}
\caption{Preprocessing of the ISIC 2019 dataset. Black image borders are detected and removed. The top row shows images of the original training data set, shown below are preprocessed images}
\label{blackedges}
\end{figure}

Training and test data of the ISIC-2019 dataset have been preprocessed to remove black areas surrounding dermoscopic images, and subsequently rescaled maintaining aspect ratio (Figure \ref{blackedges}). Descriptive text appended to images in the SD-198 dataset has been removed.

\subsection{Data Augmentation}
\begin{figure}[!t]
\centering
\includegraphics[width=3.49in]{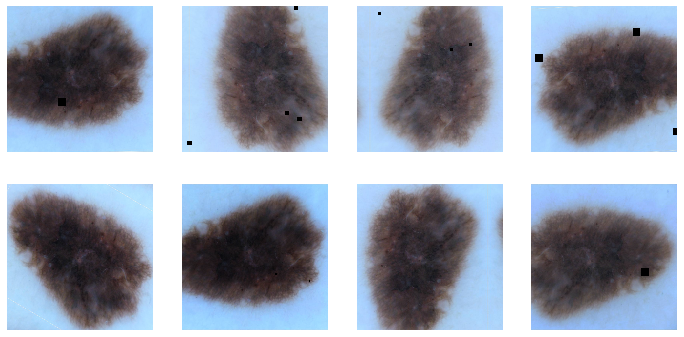}
\caption{Applied augmentations for a single training image. Random rotation, translation in the x and y directions as well as scaling within defined limits avoid overfitting on the training data and enable a better generalization of the model. Used augmentation parameters are: max\_rotate=45, p\_affine=0.5, do\_flip=True, flip\_vert=True, max\_zoom=1.05, max\_lighting=0.2, crop\_pad(input\_size), cutout(n\_holes=(1,1), length=(16,16), p=.5).}
\label{augmentations}
\end{figure}

To avoid overfitting \cite{hinton2012improving} in neural networks, dropout \cite{srivastava2014dropout} is often used. Another simple method for regularization (and expansion of the number of different training samples) of CNNs is data augmentation. During training, input data is changed randomly according to certain criteria (translation, rotation, scaling, etc.). Additionally, Cutout \cite{devries2017improved} has been used for regularization. Figure ~\ref{augmentations} shows the applied augmentations.

\subsection{Out of Distribution Detection} \label{subsec_ood}
Neural networks offer little or no guarantee of reliable prediction when applied to data that was not generated through the same process that was used to create the network's training data. With such Out-of-Distribution (OOD) inputs, the prediction may not only be incorrect but also associated with a high level of confidence \cite{goodfellow2014explaining, nguyen2015deep} of the network, which restricts the reliability of deep learning classifiers in real world applications. Often the predictions of (ensembles of) classifiers that have been trained on data within the distribution are examined for the presence of OOD inputs using statistical methods \cite{hendrycks2016baseline, lakshminarayanan2017simple}. Alternatively, the input distribution can be modeled directly by using generative models that do not require the presence of class labels. However, it has been shown that this method can also output higher probabilities on OOD inputs than on inputs within the distribution \cite{ren2019likelihood}. In the ISIC 2019 Challenge, classes that are not included in the training data set should be detected as OOD and recognized as class "UNK". In this work, a data-driven approach to the recognition of OOD inputs is pursued by using images (mostly from SD-198, see subsection ~\ref{subsec_sd198}) as training data for the "UNK" class that are not labeled as one of the classes of the ISIC-2019 training data set. However, this approach is far from optimal, and OOD detection in deep learning classifiers remaining an unsolved problem. Further work is needed to improve classifier performance regarding OOD detection.

\subsection{Dataset Imbalance}
A common problem with deep learning-based applications is the fact that some classes have a significantly higher number of samples in the training set than other classes. This difference is known as class imbalance. There are many examples in areas such as computer vision \cite{van2018inaturalist, xiao2010sun, johnson2013hybrid, kubat1998machine, beijbom2012automated}, medical diagnosis \cite{grzymala2004approach, mac2002problem}, fraud detection \cite{philip1998toward}, and others \cite{radivojac2004classification, cardie1997improving, haixiang2017learning} where this problem is highly significant and the incidence of one class (e.g. cancer) can be 1000 times less than another class (e.g. healthy patient) \cite{buda2018systematic}. It has been shown that a class imbalance in training data sets can have a significant adverse effect on the training of traditional classifiers \cite{japkowicz2002class}, including classic neural networks or multilayer perceptrons \cite{mazurowski2008training}. The class imbalance influences both the convergence of neural networks during the training phase and the generalization of a model to real or test data \cite{buda2018systematic}.

\subsubsection{Undersampling / Oversampling}
Undersampling and oversampling in data analysis are techniques to adjust the class distribution of a data set (i.e. the relationship between the different classes/categories represented). These terms are used in statistical sampling, survey design methodology, and machine learning. The goal of undersampling and oversampling is to create a balanced data set. Many machine learning techniques, such as neural networks create more reliable predictions when trained on balanced data. Oversampling is generally used more often than undersampling. The reasons for using undersampling are mainly practical and often resource-dependent. With random oversampling, the training data is supplemented by multiple copies of samples from minority classes. This is one of the earliest methods proposed that has also proven robust \cite{ling1998data}. Instead of duplicating minority class samples, some of them can be chosen at random by substitution. Other methods of handling unbalanced data sets such as synthetic oversampling \cite{chawla2002smote} are more suitable for traditional machine learning tasks \cite{fernandez2018smote} and were therefore not considered any further in this work.

\subsubsection{Weighted Cross-Entropy Loss}
Weighted cross-entropy \cite{ronneberger2015u} is useful for training neural networks on unbalanced data sets. \cite{vyas2018out} suggest adding a margin-based loss value to the cross-entropy on in-distribution training patterns in order to ensure a minimum difference in average entropy between in-distribution and out-of-distribution data. This ensemble-based method is intended to surpass previous methods of recognizing out-of-distribution inputs such as ODIN \cite{liang2017enhancing}. Cross entropy can be described as

\small $$  L(x,y) = -log\left(\frac{\exp(x[y])}{\sum_{j}\exp(x[j])}\right)
= -x[y] + log\left(\underset{j}{\sum}\exp(x[j])\right)  $$
\normalsize or, by using class weights:
\small $$ L(x,y)=W[y]\left(-x[y]+log\left(\underset{j}{\sum}\exp(x[j])\right)\right) $$
\normalsize

The arithmetic mean of the loss values achieved is calculated for each mini-batch. A weight vector can be calculated using effective class weights \cite{cui2019class} with the simple formula
$ (1- \beta ^{n}) / (1- \beta) $,
with the hyperparameter beta equal to 0.999 (a choice of the parameter beta equal to zero would not apply any weighting and a choice of beta equal to 1 would correspond to weighting by the inverse class frequency). In the simplest case, loss values can be weighted by multiplying by inverse class frequencies.

\subsubsection{Thresholding}
Also referred to as threshold shifting or rescaling, thresholding adapts the decision threshold of a classifier. This method is used at inference time and involves changing the output class probabilities. There are several ways in which the network outputs can be rescaled. In general, an optimization algorithm can be used to configure the network to minimize any criteria \cite{lawrence1998neural}. The simplest method only compensates for a priori class probabilities \cite{richard1991neural}. It has been shown that neural networks estimate Bayesian a posteriori probabilities \cite{richard1991neural}. That is, for a given data point x, the output for class c is implicitly
$ y_{i}(x)=p(c|x)=\frac{p(c)p(x|c)}{p(x)} $.
The actual probabilities of class membership can therefore be calculated by dividing the output of the network by the estimated a priori probability
$  p(c)=\frac{|c|}{\sum_{k}|k|} $,
where $|c|$ is the number of samples of class $c$ \cite{buda2018systematic}. The resulting class probabilities are normalized after thresholding is applied. This simple method of handling an existing class imbalance can significantly increase the class probability distribution approximation made by classifiers.

\subsection{Transfer Learning}
Transfer learning in the context of machine learning is a technique that uses information obtained from solving a problem and applies it to a similar problem. When using transfer learning, a model that has already been trained on another data set is adapted to custom data. Ideally, the pre-trained model has been trained on similar data, but this is not strictly necessary. The final layers of the network are removed and replaced by output layers featuring  appropriate dimensions. The model is then trained on custom data. By using transfer learning, the time required for training a network can be greatly reduced \cite{pan2008transfer, pan2010survey, hoo2016deep}. The existing pre-trained model thus serves as a feature extractor, which forwards features such as edges, texture, position of recognized objects, etc. to the last layer for classification. A softmax function (normalized exponential function) transforms the network output into a vector of numbers between zero and one which sum up to one which allows interpreting the output of the network as a probability distribution.

\subsection{Test Time Augmentation}
Data augmentation is a technique widely used to improve neural network training performance and reduce generalization errors. The same image data augmentation technique can also be used at inference time to allow the model to make predictions for several different versions of each image in the test data. Test Time Augmentation (TTA) predictions are formed by calculating the average of the regular predictions (with a weighting of beta=0.4) with the average of the predictions obtained by predicting on augmented versions of the image data (with a weighting of 1-beta). The transformations specified for the training set are applied with the following changes: Scaling with a factor of 1.05 controls the scaling for the zoom (which is not random for TTA). Furthermore, the cropping is not random to ensure that the four corners of the picture are used. Reflection is not random but is applied once to each of these corner images (so that a total of 8 augmented versions are created).

\subsection{Ensembling}
Ensembling is the use of several independently trained models to form an overall prediction. The basic idea of ensembling is that individual models have weaknesses in different areas, which are compensated by the combination with predictions of other independently trained models. Possible ensembling strategies are e.g. majority voting, the use of a weighted average based on classifier confidences, or simply using the arithmetic mean of several predictions of different models and model architectures \cite{kowsari2018rmdl}.

\section{Experiments}
The CNN architectures Inception-ResNet-v2 \cite{DBLP:journals/corr/SzegedyIV16}, SE-ResNeXt-101 (32x4d) \cite{hu2018squeeze}, NASNet-A-Large \cite{DBLP:journals/corr/abs-1712-00559}, EfficientNet-B4 and EfficientNet-B5 \cite{ DBLP:journals/corr/abs-1905-11946} pre-trained on the ImageNet data set were adapted for the task of classifying the nine classes of the ISIC-2019 Challenge by replacing final layers with a custom linear layer to output nine class probabilities. Real-time data augmentation has been used to improve the generalizability of the resulting models. Models have been trained on an NVIDIA GTX 1070 GPU. Batch sizes (number of training samples that are used for a single forward pass) were adapted to individual architectures and input sizes to achieve optimal utilization of the available video memory. Images have been resized to fit model input sizes prior training.

Models have been trained via transfer learning over 32 epochs followed by model fine-tuning using differential learning rates until convergence using One Cycle Policy \cite{smith2018disciplined}, allowing very rapid convergence rates of trained networks \cite{smith2019super}. Appropriate learning rates were determined manually at regular intervals. The use of a weighted loss function has, contrary to expectations, only proven to be advantageous for training the NASNet-A-Large architecture, which has been unable to converge without applying weighted loss. Other architectures could not benefit from training using a weighted loss function. Early stopping has been applied to avoid model overfitting. Best models have been selected based on their performance on the validation data. Out-of-distribution detection using  thresholding proved to provide inferior results to using a data-driven approach as described in ~\ref{subsec_ood}.

The unsatisfactory balanced multiclass accuracy of the NASNet model may be caused by the relatively small batch size, which was limited to four due to the size of the model. As expected, improved performance of deep neural networks in the classification of ImageNet data can be directly translated to models trained on custom data sets. Improved CNN architectures, which achieve higher accuracy in the classification of the ImageNet data set, thus also provide better results in the classification of dermoscopic images.

A rescaling of the outputs of the models by multiplying the output probabilities by inverse class frequency have proven to be advantageous for the  balanced multiclass accuracy of the network predictions in all cases where no weighted loss function has been used. Applying rescaling on models trained using a weighted loss function did not improve balanced multiclass prediction accuracy. The outputs of several independently trained models were combined into an overall prediction using the arithmetic mean of all model predictions and transmitted to the automated evaluation system of the ISIC-2019 Challenge.

Table ~\ref{table_metrics_models} shows results for individual models. Best performing models were used to form ensemble predictions. NASNet-A-Large was not included in the ensemble due to the unsatisfactory overall accuracy achieved. Although EfficientNet  shows the best results of all trained network architectures, the combination with predictions from SE-ResNeXt-101 (32x4d) and Inception-ResNet-v2 models still lead to higher average accuracy than any single model could achieve independently.

\begin{table}
%\renewcommand{\arraystretch}{1.3}
% if using array.sty, it might be a good idea to tweak the value of
% \extrarowheight as needed to properly center the text within the cells
\caption{Single Model, Ensemble Balanced Accuracy}
\label{table_metrics_models}
\centering
%% Some packages, such as MDW tools, offer better commands for making tables
%% than the plain LaTeX2e tabular which is used here.
\begin{tabular}{c | c}
Architecture & Accuracy \\
\hline
EfficientNet-B5 & 0.600 \\
SE-ResNeXt-101(32x4d) &	0.582 \\
EfficientNet-B4	& 0.577\\
Inception-ResNet-v2	& 0.569\\
NASNet-A-Large & 0.504\\
\hline
Ensemble (excluding NasNet) & 0.634\\
\end{tabular}
\end{table}

\begin{table}
%\renewcommand{\arraystretch}{1.3}
% if using array.sty, it might be a good idea to tweak the value of
% \extrarowheight as needed to properly center the text within the cells
\caption{Metrics (Ensemble)}
\label{table_metrics}
\centering
%% Some packages, such as MDW tools, offer better commands for making tables
%% than the plain LaTeX2e tabular which is used here.
\tabcolsep=0.052cm
\begin{tabular}{cc|ccccccccc}
Category & Mean \\
Metrics & Value & MEL & NV & BCC & AK & BKL & DF & VASC & SCC & UNK \\
\hline
AUC & .902 & .924 &	.957 &	.942 & .917 & .893 &	.977 &	.932 &	.936 &	.638 \\
AUC, Sens\textgreater80\% & .813 &	.853 &	.926 &	.883 &	.829 &	.776 &	.966 &	.868 &	.876 &	.336 \\
Avg. Precision
&.561 	&.766 	&.923 	&.719 	&.366 	&.572 	&.586 	&.502 	&.326 	&.285\\
Accuracy 	&.923 	&.899 	&.894 	&.908 	&.933 	&.933 	&.983 	&.978 	&.969 	&.808\\
Sensitivity
&.525 	&.581 	&.752 	&.666 	&.580 	&.384 	&.744 	&.614 	&.408 	& .00\\
Specificity 	&.973 	&.963 	&.962 	&.944 	&.952 	&.985 	&.986 	&.983 	&.982 &	1.00\\
Dice Coeff
&.491 	&.659 	&.821 	&.654 	&.468 	&.499 	&.523 	&.434 	&.364 	&\ .00\\
PPV
&.609 	&.760 	&.905 	&.642 	&.392 	&.713 	&.404 	&.335 	&.328 	& 1.00\\
NPV
&.941 	&.919 	&.890 	&.950 	&.977 	&.944 	&.997 	&.995 	&.987 	&.808 \\
\end{tabular}
\end{table}

\begin{figure}[t!]
\centering
\includegraphics[width=3.2in]{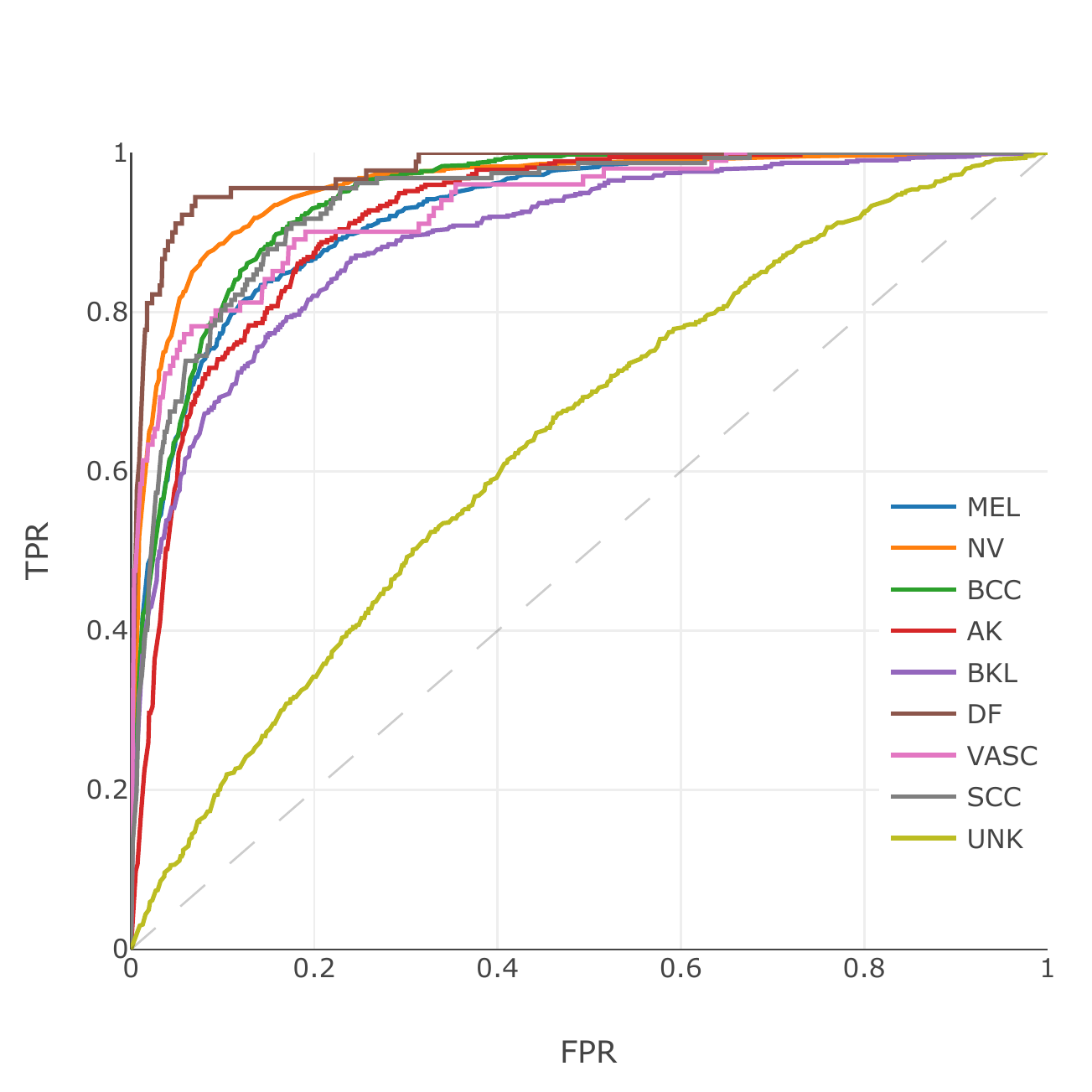}
\caption{ROC curve for the  0.634 balanced multiclass accuracy ensemble. The ROC curve shows the diagnostic capability of a binary classifier as its decision threshold varies. The ROC curve is constructed by plotting the true positive rate (TPR) against the false positive rate (FPR) at various threshold settings. The true positive rate is also referred to as sensitivity, recall or detection probability, whereas FPR corresponds to the false positive rate (1 - specificity).}
\label{roc}
\end{figure}

Table ~\ref{table_metrics} shows metrics for the ensemble with 0.634 balanced multiclass accuracy, as computed by the ISIC challenge website. AUC: Area under the receiver operating characteristic (ROC) curve. AUC, Sens \textgreater 80\%: area under the ROC curve, evaluated exclusively for the region in which the sensitivity is greater than 80\%. Average precision (precision is also called Positive Predictive Value - PPV) measures the area under the interpolated precision-recall curve (recall = sensitivity). Accuracy measures the overall accuracy of the classifier, i.e.
$ Accuracy = sensitivity * prevalence + specificity * (1-prevalence) $.
Sensitivity measures true-positive predictions, specificity (recall) measures true-negative predictions of the classifier. The F1 score (Dice Coefficient) is the harmonious mean of precision and recall, with an F1 score reaching its best value at 1 (perfect precision and recall). F1 score is also known as the Sørensen-Dice coefficient or Dice similarity coefficient (DSC). A positive predictive value (PPV) is the likelihood that subjects who test positive will actually have the disease. A negative predictive value (NPV) is the likelihood that subjects who test negative really do not have the disease. Figure ~\ref{roc} shows the receiver operating characteristic curve for the ensemble.

\section{Conclusion}
Deep learning has become a mature technology for the classification of image content and can achieve similar or superior accuracy as human experts in the classification of skin lesions. The use of deep learning applications that automatically evaluate clinical and dermoscopic images and classify skin lesions offer great potential for improving and implementing prevention and screening measures and increasing their efficiency.
One of the main criticisms of deep learning applications, that these networks have to be treated as a black box and that there is no easy explanation of how they form their decisions remain unchanged despite some progress in the visualization of network activations. Careful validation of trained models using real-world data sets before and also during use is essential.
Progress in the development of more efficient architectures of deep neural networks and improved accuracy in the classification of images with high image quality does not automatically mean that results can be transferred to real-world applications. For instance, \cite{beede2020human} examined the use of a classification system created by Google researchers to detect diabetic retinopathy in 11 clinics in Thailand and found that this technology does not yet work well in practice despite all the research advances.
Advantages of deep learning applications in the medical field are the rapid availability of diagnosis compared to analysis by human specialists and cost-effective provisioning of models for large numbers of simultaneous users. Central provisioning of deep learning models allows uncomplicated and transparent delivery of improved models without having to make changes to client software. Cloud applications can serve current deep learning models cost-effectively through automatic horizontal scaling of active services and flexible price calculations. Also, deep learning applications can help nursing staff to better argument their own assessments to specialists and to prioritize urgent cases accordingly. Even if decisions made by deep learning models still have to be manually verified by human experts, automated image classifiers can support these human experts and reduce the workload by accelerating decision making processes, therefore contributing to more efficient utilization of the resources of health systems.

% or


%\appendices

%\section{}

%\section*{Acknowledgment}


\ifCLASSOPTIONcaptionsoff
  \newpage
\fi



%\bibliographystyle{IEEEtran}
%\bibliography{IEEEtran/bib/references.bib}

% https://stackoverflow.com/a/11884947
\bibliographystyle{IEEEtran}
\nocite{*}
\bibliography{references.bib}

\end{document}